\documentstyle[12pt,epsf]{article}
\begin{document}

\title{Gauge independent effective gauge fields}

\author{~Kirill~A.~Kazakov\thanks{Email address: $kirill@theor.phys.msu.su$} ~and
~Petr~I.~Pronin\thanks{Email address: $petr@theor.phys.msu.su$}}

\maketitle

\begin{center}
{\em Moscow State University, Physics Faculty,\\
Department of Theoretical Physics.\\
$117234$, Moscow, Russian Federation}
\end{center}

\begin{abstract}
The problem of gauge independent definition of the effective gauge field
is considered. The Slavnov identities corresponding to a system of
interacting quantum gauge and classical matter fields, playing the role
of a measuring device, are obtained. With their help, in the case of
power-counting renormalizable theories, gauge independence of the
effective device action is proved in the low-energy limit, which
allows to introduce the gauge independent notion of the effective gauge field.
\end{abstract}

\unitlength=1mm


\unitlength=1pt

\section{Introduction}

Description of quantized fields by means of the effective action (EA)
is the most general in quantum theory. Being the sum of all
one-particle-irreducible diagrams EA for a given theory allows to
calculate any Green function in this theory. Well known that
it also can be given a nonperturbative definition as the Legendre
transform of the Green functions generating functional logarithm.
Formal analogy between the classical equations of motion and
quantum equations describing dynamics of the mean fields suggests
natural interpretation of EA as the quantum substitute for its classical
counterpart. However, explicit dependence of EA on the way the theory
is quantized lacks direct physical application of this remarkable
analogy.
The most important kind  of such a dependence, which attracts our attention
in this Paper, is the gauge dependence of EA for gauge theories.

It is not our purpose to anew investigate various procedures formulated
by many authors in attempts to construct a gauge independent object
from EA. Instead, we would like to pay attention to a possible {\it physical}
reason for the gauge dependence of EA, recently pointed out by Dalvit and
Mazzitelli \cite{dalvit}. In the case of quantum gravity
they showed that the motion of a classical device measuring the effective
gravitational field is independent of the choice of gauge conditions
fixing the general coordinate invariance. More precisely, the equations
of motion (geodesic equation) of a test particle in the effective static
gravitational field of a point mass, calculated in the one-loop
approximation up to leading logarithms was shown to be independent
of the choice of linear gauge.

The point is that while the graviton-test particle quantum interaction
is negligible in calculation of the total effective gravitational field,
it is {\it not} when the equations of the test particle motion are
to be determined. It turns out that in the latter case the gauge
dependent part of the contribution due to graviton-test particle
interaction just cancels that corresponding to the ordinary gauge
dependence of the mean field.

This fact offers a tempting possibility to change our plain view on the problem of
gauge independent definition of the effective gravitational field,
and look at it through a prism of the measurement. In other
words, we can try to describe the effective gravitational field
in terms corresponding to the measuring device. For example,
in the case considered in \cite{dalvit} it is the form of the equations
of the test particle motion by which the effective gravitational field
is implicitly described.

Whether a proper definition of the effective field can be given in this way,
depends on resolution of the following questions:
\begin{enumerate}
\item Whether the special choices of the source for the gravitational field
and of the measuring device made in \cite{dalvit} are essential for
the aforementioned cancellation.
\item Whether this
cancellation holds at any order of the loop expansion and for all
energies (not only for the one-loop low-energy leading quantum corrections).
\item If the effective gravitational field is described through
characteristics of the measuring device, is such a description
actually independent of the choice of device, for the concept
of the effective action to be self-contained.
\item Is all of this inherent to the
gravitation, or represents a general property of gauge interactions.
\end{enumerate}

The purpose of this Paper is to show that the answer to 1.,4., and to
the first part of 2. is really positive, i.e. the low-energy
leading quantum corrections to the equations of motion of any kind
of classical matter (infinitely weak) interacting with the gravitational or
any other gauge field are gauge independent at any order of the loop expansion.
In sec.2 we introduce notations and display some basic tools used later
in investigation of EA properties. In sec.3 the Slavnov identities
for the generating functionals of the Green functions corresponding
to the system gauge field plus device are derived, on which basis
the renormalization equations for divergent parts of these functionals
are obtained in sec.4. These equations allow to demonstrate the gauge
dependence cancellation most generally. In sec.5 we briefly discuss
the rest of the problems listed above, and make conclusions.

\section{The quantum effective action}

The reason for the cancellation of the gauge dependence found by Dalvit and Mazzitelli
may lie, of course, only in the residual symmetry of the Faddev-Popov
quantum action for the gauge field, the Becci-Rouet-Stora-Tyutin (BRST) symmetry \cite{brst}.
Having the form of the ordinary gauge transformation for the gauge and matter
fields, the latter is indifferent to the specific structure of the classical
action for these fields. Therefore, following the standard procedure of derivation
of the Slavnov identities for the generating functionals of the Green functions,
we can try to obtain analogous identities for the system of the gauge field
plus measuring device in the most general form.

We consider a general type gauge
theory described by an action $S(A_{a},\phi_{i})$, where $A_{a}, a = 1,...,n$
denotes the gauge field and $\phi_{i}, i = 1,...,m$ -- matter fields of any kind.

If the pure gauge theory describes free fields $A_{a},$ then
a number of {\it quantum} matter fields interacting with $A_{a}$ should be
included in $\phi$. However, for notation simplicity we suppose that the
gauge field is self-interacting and $\phi$ contains only {\it classical}
matter fields. Furthermore, the part of $\phi$ corresponding to the sources
for $A_{a}$ can be omitted since any desired A-field configuration
can be formally obtained by appropriate choice of the standard source term
$J_{a} A_{a}$ which is normally introduced into the generating
functional of the Green functions. Thus, we suppose the fields $\phi_{i}$ to
describe the measuring device only. The latter is a classical object in
the ordinary sense that the {\it low-energy} quantum corrections to its equations of
motion due to propagation of the $\phi$-fields can be neglected, which usually means that the device should be
sufficiently heavy.
Following \cite{dalvit} we also require the device
contribution to the total gauge field to be infinitely small.
One could suppose, for example, that the coupling constants of
the gauge field-device interaction are sufficiently small.
Since, however, it is not always possible to choose these
constants arbitrary small\footnote{In the case of non-abelian gauge theories
their value may be fixed already by the form of gauge field self-interaction.},
we simply imagine that the device action enters the full action
with a small overall coefficient.

It is problematical to satisfy the above
requirements in the case of gravity, since they contradict to each other.
Even more: in this case we cannot satisfy the first of them alone because
there is no such a thing as the classical source for gravity, as was
pointed out in \cite{donoghue}. Therefore, in the case of measurement of the
effective gravitational field we are forced to introduce the classical
form for the device action "by hands".

Let the action $S(A,\phi)$ be invariant under the following
(infinitesimal) gauge transformations
\begin{eqnarray}&&
\delta A_{a} = D_{a\alpha}(A)\xi_{\alpha},
~~\delta\phi_{i} = \tilde{D}_{i\alpha}(\phi)\xi_{\alpha},
\end{eqnarray}
where $D_{a\alpha}(A), \tilde{D}_{i\alpha}(\phi)$ are the generators,
and $\xi_{\alpha}, \alpha = 1,...,N$ are arbitrary gauge functions of the gauge transformations.
We suppose that these generators form a closed algebra
\begin{eqnarray}&&
D_{a\alpha,b} D_{b\beta} - D_{a\beta,b} D_{b\alpha} = f_{\alpha\beta\gamma} D_{a\gamma},
\nonumber\\&&
\tilde{D}_{i\alpha,j} \tilde{D}_{j\beta} - \tilde{D}_{i\beta,j} \tilde{D}_{j\alpha} = f_{\alpha\beta\gamma} \tilde{D}_{i\gamma},
\end{eqnarray}
where the "structure constants" $f_{\gamma\alpha\beta}$ are some linear differential
operators which we assume to be field-independent, for simplicity. Commas
followed by indices denote functional differentiation with respect
to the corresponding fields, and DeWitt's summation-integration on repeated
indices is supposed.

To fix this invariance we impose an arbitrary gauge condition $F_{\alpha}(A) = 0$.
For simplicity, we suppose that it is linear in the field $A$: $F_{\alpha}(A) \equiv F_{\alpha,a} A_{a}$,
where $F_{\alpha,a}$ is some (differential) operator independent of the fields.
Weighted in the usual way this gauge condition enters the Faddeev-Popov (FP)
quantum action $S_{fp}$ in the form of the gauge-fixing term
\begin{eqnarray}
S_{gf} = - \frac{1}{2\xi}F_{\alpha}^{2},
\end{eqnarray}
$\xi$ being the weighting parameter. Introducing FP ghost fields $C_{\alpha}, \bar{C}_{\alpha}$
we write the FP quantum action
\begin{eqnarray}
S_{fp} = S(A,\phi) + S_{gf} + \bar{C}_{\beta}F_{\beta,a}D_{a\alpha}C_{\alpha}.
\end{eqnarray}
$S_{fp}$ is still invariant under the following BRST transformations \cite{brst}
\begin{eqnarray}&&\label{brst}
\delta_{brst}A_{a} = D_{a\alpha}(A)C_{\alpha}\lambda,
\nonumber\\&&
\delta_{brst}\phi_{i} = \tilde{D}_{i\alpha}(\phi)C_{\alpha}\lambda,
\nonumber\\&&
\delta_{brst}C_{\alpha} = - f_{\alpha\beta\gamma}C_{\beta}C_{\gamma}\lambda,
\nonumber\\&&
\delta_{brst}\bar{C}_{\alpha} = - \frac{1}{\xi}F_{\alpha}\lambda,
\end{eqnarray}
$\lambda$ being a constant anticommuting parameter.

To be able to write down the Slavnov identities for the generating
functional of connected Green functions we introduce sources
for the BRST transformations following Zinn-Justin \cite{zinnjustin}, and obtain the quantum
action $\Sigma$ in the form
\begin{eqnarray}\label{qaction}
\Sigma = S(A,\phi) - \frac{1}{2\xi}F_{\alpha}^{2} + \bar{C}_{\alpha}F_{\alpha,a}D_{a\beta}C_{\beta} + K_{a}D_{a\alpha}C_{\alpha} + \tilde{K}_{i}\tilde{D}_{i\alpha}C_{\alpha}.
\end{eqnarray}
Introduction of the source $\tilde{K}$ is dispensable since $\tilde{D}_{i\alpha}C_{\alpha}$
is linear in quantum fields. However, it allows to write the Slavnov
identities in the form containing no explicit information on the structure
of the gauge algebra, and in addition to that, to omit all ghost sources
from the very beginning  (these sources should be restored, however, in
renormalization of the theory).

Below we consider the most important kind of the gauge dependence
of EA, namely its dependence on the weighting parameter
$\xi$. The general case contains no principal complications, and can be
handled, for example, by extending the field content of the theory to
include a number of auxiliary fields introducing the gauge, and employing
the method of anticanonical transformations ( see, e.g., \cite{lavtut}).
A natural way of investigation of the $\xi$-dependence is to introduce
the term
\begin{eqnarray}
Y F_{\alpha}\bar{C}_{\alpha},
\end{eqnarray}
$Y$ being a constant anticommuting parameter,
into the quantum action \cite{kluberg}. Thus we write the generating
functional of the Green functions as\footnote{We suppose that all divergent
quantities appearing below are invariantly regularized. For technical
simplicity we assume also that $\delta (0) = 0$ in this regularization,
to omit possible local factors in the functional integral measure.}
\begin{eqnarray}&&\label{gener}
Z[J,\phi,K,\tilde{K},Y] =
{\displaystyle\int}dA dC d\bar{C} exp\{i (\Sigma(A,\phi,C,\bar{C},K,\tilde{K})
+ Y F_{\alpha}\bar{C}_{\alpha} + J_{a} A_{a})\}.
\end{eqnarray}
$\phi$-fields are not integrated in Eq. (\ref{gener}). Following \cite{dalvit}
we consider them as c-functions, and the absence of these fields in the
integral measure reflects the fact that we neglect all the quantum
]contributions due to their propagation.

\begin{figure}
\epsfxsize=16cm\epsfbox{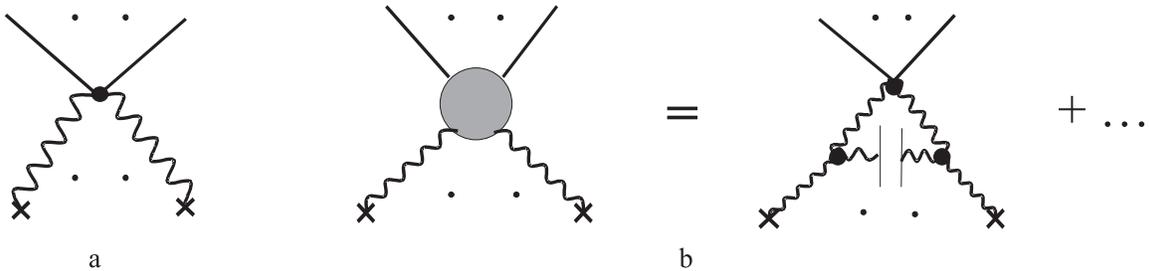}
\caption{A typical vertex representing local interaction of the measuring
device (straight lines) with the effective gauge field (wavy lines) in
the standard interpretation (a). The corresponding non local vertex (the
shaded blob) given by the generating functional (8) (b).
The plus sign denotes higher-order contributions. The source $J$ for the
gauge field is pictured schematically as the local crossed vertex.}
\end{figure}

Diagrammatically, this situation is illustrated in Fig.1.
Fig.1(a) represents a typical vertex of the gauge field-device interaction
according to the standard definition of the effective field as the quantum
average of the corresponding field operator. It is implied by this
definition that the mean field is simply put into the classical equations
of device motion instead of its tree value. Such vertices are local, unlike
those given by the generating functional (\ref{gener}) and represented in Fig.1(b).
To determine effective equations of device motion
we have to consider the sum of diagrams like that pictured in Fig.1(b),
each having only one insertion of a $\phi$-vertex, since the
device action is supposed to be infinitely small.

Now, introducing the generating functional of the connected Green functions
\begin{eqnarray}
W(J) = - i~ln~Z(J),
\end{eqnarray}
we define the effective action $\Gamma$ as the Legendre transform of $W$
with respect to the mean gauge field
\begin{eqnarray}\label{mean}
A_{a} = \frac{\delta W}{\delta J_{a}}
\end{eqnarray}
(denoted by the same symbol as the corresponding field operator):
\begin{eqnarray}
\left.\Gamma (A) = W(J) - J_{a} A_{a}\right|_{J \to J(A)},
\end{eqnarray}
where the function $J(A)$ is implicitly defined by Eq. (\ref{mean}).

In the standard interpretation the reciprocal of Eq. (\ref{mean})
\begin{eqnarray}\label{meanrcp}
\frac{\delta\Gamma}{\delta A_{a}} = - J_{a}
\end{eqnarray}
are the effective equations of motion for the full quantum corrected field
$A$ corresponding to the given background field configuration $A^{0}$ satisfying
\begin{eqnarray}\label{tree}
\frac{\delta S}{\delta A_{a}}(A^{0}) = - J_{a}.
\end{eqnarray}
The use of $J$ as the source for the field $A^{0}$ instead of realistic
matter sources, though formal, allows to simplify the derivation of
the Slavnov identities below. As always, the source $J$ satisfies
the "conservation law"
\begin{eqnarray}
D_{a\alpha}(A^{0})J_{a} = 0,
\nonumber
\end{eqnarray}
where $A^{0}$ is the solution of Eq. (\ref{tree}), satisfying
\begin{eqnarray}\label{gauge}
F_{\alpha}(A^{0}) = 0.
\end{eqnarray}

\section{The Slavnov identities}
\subsection{Preliminaries}
Being a classical object the measuring device is completely described
by its action. We can therefore investigate the gauge dependence
of the latter rather than of the corresponding equations of motion
(as was done in \cite{dalvit}). The action for the measuring device
is the part of $\Gamma$ containing the fields $\phi_{i}$.
Its gauge dependence is determined by the Slavnov identities
 for the generating functional of proper vertices corresponding to (\ref{gener}).
However, these identities are complicated because of the peculiar role
of the device action which is a kind of source for the gauge field.
It is the nonlinearity of this source on the fields which complicates
the usual derivation of the Slavnov identities.

Fortunately, in the present case we can limit ourselves by derivation of
the Slavnov identities for the functional $W$ only.
Indeed, the gauge dependent part of the device action is a sum
of two different contributions.
The first is the ordinary explicit gauge dependence of the effective action.
The second results from implicit gauge dependence of the mean field $A$.
Being a solution of the gauge dependent effective equations (\ref{meanrcp})
the latter is also gauge dependent. It is precisely this gauge dependence
of $A$ which lacks its physical interpretation. Thus, denoting by $\Gamma_{\phi}$ the
part of $\Gamma$ containing $\phi$-fields we have for the full variation
of the device action under a small change $\delta\xi$ of the gauge parameter $\xi$:
\begin{eqnarray}\label{full}
d\Gamma_{\phi} = \frac{\partial\Gamma_{\phi}}{\partial\xi}\delta\xi
+ \left.\frac{\delta\Gamma_{\phi}}{\delta A_{a}}\frac{\partial A_{a}}{\partial\xi}\right|_{J}\delta\xi \equiv \frac{d\Gamma_{\phi}}{d\xi}\delta\xi.
\end{eqnarray}
In (\ref{full}) the derivative $\partial A_{a}/\partial\xi$ is calculated
keeping $J$ fixed in accordance with the meaning of $J$ as producing the given
classical field $A^{0}.$

Now note, that if we define the quantity $W_{\phi}$ by analogy with $\Gamma_{\phi}$,
i.e., as the part of $W$ containing $\phi$, then
\begin{eqnarray}\label{equiv}
\Gamma_{\phi}(A) = \left.W_{\phi}(J)\right|_{J \to J(A)},
\end{eqnarray}
since the device action is supposed to be infinitely small.

Comparing (\ref{full}) and (\ref{equiv}) we arrive at the following important
relation
\begin{eqnarray}\label{main}
\frac{d\Gamma_{\phi}(A,\xi)}{d\xi} = \left.\frac{\partial W_{\phi}(J,\xi)}{\partial\xi}\right|_{J \to J(A)}.
\end{eqnarray}

Thus, perhaps needed in carrying out the renormalization program, the Slavnov
identities for $\Gamma$ turn out to be unnecessary in our consideration.

Let us now go over to the successive derivation of the Slavnov identities for $W.$

\subsection{Derivation}
Following the standard procedure (see, e.g., \cite{zinnjustin}) we perform
a BRST shift (\ref{brst}) of integration variables in (\ref{gener}).
Unlike the usual case, however, the quantum action (\ref{qaction}) is not invariant
under this operation, since besides the quantum fields $A, C, \bar{C}$ it contains
the classical field $\phi$ which is not integrated in (\ref{gener}).
Therefore, we obtain the following identity\footnote{The corresponding Jacobian $\sim exp\{\delta (0) ... \} = 1.$}
\begin{eqnarray}&&\label{slav}
\hspace{-0,5cm}{\displaystyle\int}dA dC d\bar{C} \left[\frac{\delta S_{\phi}}{\delta A_{a}} D_{a\alpha} C_{\alpha}
+ J_{a} D_{a\alpha}C_{\alpha} + Y \bar{C}_{\alpha} F_{\alpha,a} D_{a\alpha} C_{\alpha}
- \tilde{K}_{i} \tilde{D}_{i\alpha} f_{\alpha\beta\gamma} C_{\beta} C_{\gamma} - \frac{Y}{\xi} F_{\alpha}^{2} \right]\times
\nonumber\\&&
exp\{i (\Sigma(A,\phi,C,\bar{C},K,\tilde{K})
+ Y F_{\alpha}\bar{C}_{\alpha} + J_{a} A_{a})\} = 0,
\end{eqnarray}
where $S_{\phi}$ is the classical device action.

Since $S_{\phi}(A,\phi)$ is invariant under BRST transformations (\ref{brst}),
we may write
\begin{eqnarray}&&
\frac{\delta S_{\phi}}{\delta A_{a}} D_{a\alpha}(A) C_{\alpha} = - \frac{\delta S_{\phi}}{\delta\phi_{i}}\tilde{D}_{i\alpha}(\phi)C_{\alpha}.
\end{eqnarray}
Then the first term in square brackets in the left hand side of (\ref{slav}) can
be transformed as
\begin{eqnarray}&&
{\displaystyle\int}dA dC d\bar{C}\frac{\delta S_{\phi}}{\delta A_{a}} D_{a\alpha}(A) C_{\alpha} exp\{...\} =
- {\displaystyle\int}dA dC d\bar{C}\frac{\delta S_{\phi}}{\delta\phi_{i}} \tilde{D}_{i\alpha}(\phi) C_{\alpha} exp\{...\}
\nonumber\\&&
\hspace{-0,8cm} = - \frac{1}{i}\frac{\delta}{\delta\phi_{i}}{\displaystyle\int}dA dC d\bar{C} \tilde{D}_{i\alpha}(\phi) C_{\alpha} exp\{...\} +
{\displaystyle\int}dA dC d\bar{C} \tilde{D}_{i\alpha}(\phi) C_{\alpha} \tilde{K}_{l}\frac{\delta\tilde{D}_{l\beta}(\phi)}{\delta\phi_{i}}C_{\beta} exp\{...\}
\nonumber\\&&
= \frac{\delta^{2} Z}{\delta\phi_{i}\delta\tilde{K}_{i}} + {\displaystyle\int}dA dC d\bar{C}\tilde{K}_{i}\tilde{D}_{i\gamma}(\phi) f_{\gamma\alpha\beta} C_{\alpha} C_{\beta} exp\{...\},
\end{eqnarray}
where locality of generators $\tilde{D}(\phi)$, and the property $\delta (0) = 0$
were taken into account.
The latter also implies that the third term in square brackets in (\ref{slav})
is equal to zero. Indeed, performing a shift $\bar{C} \to \bar{C} + \delta\bar{C}$
of integration variables in the functional integral (\ref{gener}) we obtain
the quantum ghost equation of motion
\begin{eqnarray}
{\displaystyle\int}dA dC d\bar{C}\left[F_{\alpha,a} D_{a\beta}(A) C_{\beta} - Y F_{\alpha} \right] exp\{...\} = 0,
\end{eqnarray}
from which follows that\footnote{We use the property $Y^{2} = 0.$}
\begin{eqnarray}
Y {\displaystyle\int}dA dC d\bar{C}\left[\bar{C}_{\alpha} F_{\alpha,a} D_{a\beta}(A) C_{\beta} - N \delta (0) \right] exp\{...\} = 0.
\end{eqnarray}
Thus, the identity (\ref{slav}) can be rewritten as
\begin{eqnarray}\label{slav0}
\left( J_{a}\frac{\delta}{\delta K_{a}} + i \frac{\delta^{2}}{\delta\phi_{i}\delta\tilde{K}_{i}} - 2 Y\xi\frac{\partial}{\partial\xi}\right) Z  = 0.
\end{eqnarray}
This is the sought identity for the generating functional of the Green
functions. It can be called {\it effective Slavnov identity}, since it is
obtained under certain conditions concerning the device motion.
In terms of the functional $W$ it looks like
\begin{eqnarray}\label{slav1}
J_{a}\frac{\delta W}{\delta K_{a}} - \frac{\delta W}{\delta\phi_{i}}\frac{\delta W}{\delta\tilde{K}_{i}} + i \frac{\delta^{2} W}{\delta\phi_{i}\delta\tilde{K}_{i}} = 2 Y\xi\frac{\partial W}{\partial\xi}.
\end{eqnarray}
In the next section (\ref{slav1}) will be used to prove the low-energy
gauge independence of the effective device action.

\section{The gauge dependence cancellation}
\subsection{The renormalization equation}
Definition of the device as a classical object, reflected in the way its
action is introduced into the generating functional $Z,$ implies
certain conditions under which the device motion can be considered in
such a way, namely, it corresponds to the {\it effective}
description of the device motion at low energies.
Well known \cite{donoghue,eff}, that in this case the leading quantum contribution to EA
is due to non-analytical terms in the amplitudes, containing the logarithms of external momenta.
On the other hand, the form of the latter can be simply read off from divergent parts of the
amplitudes, since it is propagation of massless particles of the theory,
which dominates at low energies (see, e.g., \cite{donoghue,vil1}). For example, in the case of dimensionally regularized  Feynman
integrals of the type
\begin{eqnarray}\label{int}&&
\mu^{\varepsilon}{\displaystyle\int}d^{4-\varepsilon}q f(q,p_{1},...,p_{n}),
\end{eqnarray}
where $\varepsilon$ -- dimensional regulator, $\mu$ -- mass scale, and
$f(q,p_{1},...,p_{n})$ -- the result of all subintegrations, the low-energy
leading contributions, corresponding to some powers of the logarithms
of the external momenta $p_{1},...,p_{n}$, are given by zero order terms
in the Loran expansion for (\ref{int}) in powers of $\varepsilon$,
and unambiguously determined by the poles of (\ref{int}).

Thus, to determine the full gauge dependence of the device effective action,
it is sufficient, in view of the relation (\ref{main}), to investigate
that of the divergent parts ($W^{div}$) of the generating functional $W$.

To do this, we use the Slavnov identity (\ref{slav1}) to obtain
the renormalization equation for $W^{div}.$ Namely, we first
separate the $Y$-dependent part of $W$
$$W = W_{1} + Y W_{2},$$
and substitute it in (\ref{slav1}). Comparing multiples of $Y$ from the
left and right hand sides of this identity then gives
\begin{eqnarray}\label{slav2}
2\xi\frac{\partial W_{1}}{\partial\xi} = - J_{a}\frac{\delta W_{2}}{\delta K_{a}} + \frac{\delta W_{1}}{\delta\phi_{i}}\frac{\delta W_{2}}{\delta\tilde{K}_{i}} - i \frac{\delta^{2} W_{2}}{\delta\phi_{i}\delta\tilde{K}_{i}},
\end{eqnarray}
where all the sources except $J_{a}$ are set equal to zero after differentiation.

Next, we extract the $\phi$-dependent part of (\ref{slav2}) and obtain
the following identity
\begin{eqnarray}\label{slav3}
2\xi\frac{\partial W_{1\phi}}{\partial\xi} = - J_{a}\frac{\delta W_{2\phi}}{\delta K_{a}} + \frac{\delta W_{1\phi}}{\delta\phi_{i}}\frac{\delta W_{2\bar\phi}}{\delta\tilde{K}_{i}} - i \frac{\delta^{2} W_{2\phi}}{\delta\phi_{i}\delta\tilde{K}_{i}},
\end{eqnarray}
where the symbol $W_{2\bar\phi}$ denotes the part of $W_{2}$ independent
of the gauge field-device interaction.
All of these identities are derived for invariantly regularized, but still
unrenormalized functionals. Being connected with the high-energy behavior
of the Green functions the renormalization of EA is immaterial in
determination of the low-energy quantum corrections to the device motion.
On the other hand, since we use the formal correspondence between divergences
of EA and the form of logs in reconstruction of the leading
quantum contributions to $\Gamma_{\phi},$ the effect of renormalization
on the structure of divergences might seem to be important for us.
However, as we have mentioned above, the use of the generating functional
of the Green functions in the form of Eq. (\ref{gener}) is justified only
in the low-energy regime of the device motion. Instead, the renormalization
of the theory must be carried out, of course, in terms of the ordinary
generating functional for which Eq. (\ref{gener}) is just an effective
expression, and in which all the fields, including those corresponding
to the measuring device, are considered as quantum. Thus, at each given
order of the loop expansion it has to be supposed that all the
subdivergences of the Green functions have been eliminated at lower orders
according to the standard procedure, so that the only superficially
divergent diagrams are in rest. It is the general result of the renormalization
theory \cite{lavtut,zinnjustin} that this procedure can be arranged in the way that preserves the
symmetry properties of the generating functionals of the Green functions.
Thus, we suppose that the functional $W$ renormalized, say, up to $(n-1)$ th-loop
order, satisfies the identity (\ref{slav3})\footnote{Strictly speaking,
in derivation of the Slavnov identities for {\it renormalized}
generating functionals $Z, W, W_{\phi}$ a possible {\it implicit} gauge dependence
of the counterterms should be taken into account, which results in additional
divergent structures appearing in these identities \cite{kazakov}.
However, we omit them in the {\it effective} Slavnov identities (\ref{slav0}),
(\ref{slav1}) since these additional terms describe purely high-energy properties of the
underlying theory.} and has local divergences of order $n.$

As follows from Eq. (\ref{main}) divergences of $\partial W_{1\phi}/\partial\xi$
are to be determined after the substitution $J \to J(A)$ has been made.
As always, this means that the corresponding one-particle-irreducible\footnote{Irreducible with respect to $A$-lines.}
diagrams should be considered only. In the present case, however, one may
substitute $J \to J(A^{0})$ directly, the function $J(A^{0})$ being determined
by Eq. (\ref{tree}).
Indeed, additional divergences associated with the reexpressing of the right hand side
of Eq. (\ref{slav3}) in terms of the mean field $A$, can appear, by assumption,
only in the $n$ th-loop order. However, they actually do not contribute
at this order, since the right hand side of Eq. (\ref{slav3}) vanishes
at the zeroth order, as one can easily verify\footnote{This corresponds
to the fact that at the tree level the device action is obviously independent
of the gauge parameter $\xi$ weighting the gauge condition.}.

Thus, splitting $W_{\phi}$ into the sum of divergent and convergent parts
$$W_{\phi} = W_{\phi}^{div(n)} + W_{\phi}^{con},$$
and noting that the corresponding parts of the identity (\ref{slav3}) must
cancel independently, we obtain the {\it renormalization equation} for $W_{\phi}^{div(n)}$\footnote{The term
$- i\delta^{2} W_{2\phi}^{div(n)}/\delta\phi_{i}\delta\tilde{K}_{i}$ is omitted in Eq. (\ref{slav4}),
since it is proportional to $\delta(0)$ due to locality of divergences.}:
\begin{eqnarray}\label{slav4}&&
2\xi\frac{\partial W_{1\phi}^{div(n)}}{\partial\xi} = - J_{a}\frac{\delta W_{2\phi}^{div(n)}}{\delta K_{a}}
+ \frac{\delta W_{1\phi}^{div(n)}}{\delta\phi_{i}}\frac{\delta W_{2\bar\phi}^{(0)}}{\delta\tilde{K}_{i}}
+ \frac{\delta W_{1\phi}^{(0)}}{\delta\phi_{i}}\frac{\delta W_{2\bar\phi}^{div(n)}}{\delta\tilde{K}_{i}},
\end{eqnarray}
where the superscript $(0)$ denotes the zeroth order approximation.

Let us now turn to examination of the right hand side of Eq. (\ref{slav4}).

\subsection{The power counting}

We begin with definition of vertices and field propagators in
the loop expansion. According to the standard procedure, one expands
the exponent of the integrand in Eq. (\ref{gener}) around the extremal $A^{0}$
\begin{eqnarray}&&
\hspace{-0,6cm}\Sigma(A,\phi,C,\bar{C},K,\tilde{K}) + Y F_{\alpha}\bar{C}_{\alpha} + J_{a} A_{a} =
\Sigma(A^{0},\phi,C,\bar{C},K,\tilde{K}) + J_{a} A^{0} +
K_{a}D_{a\alpha}(A^{0}) C_{\alpha}
\nonumber\\&&
\hspace{-0,2cm} + \tilde{K}_{i}\tilde{D}_{i\alpha}(\phi) C_{\alpha} +
\frac{1}{2} \left( S + S_{gf} \right)_{,ab}(A^{0},\phi) a_{a} a_{b} + Y F_{\alpha}(a)\bar{C}_{\alpha}
+ K_{a} D_{a\alpha,b}(A^{0}) a_{b} C_{\alpha} + \cdot\cdot\cdot ~,
\nonumber
\end{eqnarray}
where the ellipsis denote terms of cubic and higher order in the quantum fields
$a \equiv A - A^{0}, ~C, ~\bar{C}.$ Note that in view of Eq. (\ref{gauge})
the term $Y F_{\alpha}(A^{0})\bar{C}_{\alpha}$ is absent in this expansion.
Therefore, the second term in the right hand side of Eq. (\ref{slav4}) vanishes identically.

For further examination of Eq. (\ref{slav4}) it is necessary to employ
the dimensional analysis.
At this point we have to limit our general consideration and require
the theory to be power-counting-renormalizable. Although quantum consequences
of the original gauge symmetry of the classical action are normally
expressed in the same form (like that of Eq. (\ref{slav4})) at all orders
of the loop expansion even despite possible deformations of the gauge algebra,
the strength  of divergences of Feynman diagrams varies from order to order, in general.
However, it is a common feature of all power-counting-renormalizable gauge theories
that the degree of divergence $D$ of an arbitrary diagram with a set
$\{n_{F}\}$ of external lines, where $\{F\} = \{A^{0},Y,K,\tilde{K}\}$, is less than or equals to
\begin{eqnarray}\label{degree}&&
\tilde{D} \equiv 4 - 2 n_{K} - 2 n_{\tilde{K}} - (2 - \sigma) n_{Y} - \sigma n_{A^{0}},
\end{eqnarray}
$\sigma$ being the canonical dimension of the gauge field $a.$
The case $D < \tilde{D}$ corresponds to theories with superrenormalizable interactions.

It can be inferred from Eq. (\ref{degree}) that in the case of $\sigma = 1$ (e.g., Yang-Mills theories)
the third term in the right hand side of Eq. (\ref{slav4}) is zero.
Indeed, the only divergent diagram with $n_{\tilde{K}} = n_{Y} = 1$
in this case corresponds to $n_{A^{0}} = 1,$ and turns into zero,
since the ghost vertex connected with the external $\tilde{K}$-line
by the ghost propagator, contains the gauge condition operator
$F_{\alpha,a}$ which vanishes upon acting
on the rest of the diagram. This is illustrated in Fig.2.

\begin{figure}
\epsfxsize=16cm\epsfbox{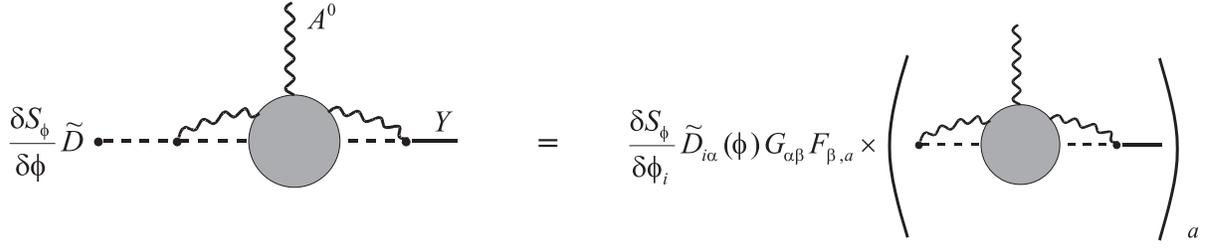}
\caption{Graphical representation of the third term in the right hand side
of Eq. (28). Dashed lines correspond to FP ghost propagator $G$.}
\end{figure}

As far as the case $\sigma = 0$ is concerned (e.g., $R^{2}$-gravity), there is an infinite
number of logarithmically divergent diagrams with  $n_{\tilde{K}} = n_{Y} = 1$
and arbitrary number of external gauge fields. In this case the above
argument goes if we confine ourselves by calculation of the gauge invariant
part of the device action only\footnote{Which is sufficient
for determination of the low-energy effective device action.}.

Finally, from (\ref{degree}) follows that if the $\phi$-vertex were absent, then
the remaining term in the right hand side of Eq. (\ref{slav4})
would diverge, with $\tilde{D} = \sigma$. Whether it does depends on the
form of the device-gauge field interaction. Obviously, insertion of a vertex
corresponding to this interaction makes a diagram with $\tilde{D} = \sigma$ convergent
if and only if
\begin{eqnarray}\label{degree1}&&
N_{\partial} + \sigma N_{a} < 4 - \sigma,
\end{eqnarray}
where  $N_{a}, N_{\partial}$ are numbers of the gauge fields entering the
vertex, and acting on them derivatives, respectively.
This condition is obviously satisfied if the full underlying quantum
theory of interacting gauge and matter fields is also power-counting
renormalizable.

Summing up, the right hand side of Eq. (\ref{slav4}) turns out to be zero,
thus proving gauge independence of the low-energy effective action
of the measuring device.

\section{Discussion and Conclusion.}

We have shown that in the case when the quantum propagation of the fields
describing measuring device can be neglected, namely, in the low-energy
classical limit, the effective equations of device motion turn out to be
gauge independent at any order of the loop expansion\footnote{Although used
in our proof, the requirement of power-counting renormalizability of the
underlying quantum theory does not seem essential for the validity of the
main result, as shows the particular example of \cite{dalvit}.}.

This allows to define in the same limit the gauge independent effective
gauge field as the field that enters these equations and couples to the
measuring device in the classical fashion.
We would like to emphasize that it is purely classical nature of the
observables (which are functionals of the $\phi$-fields) due to which
the well-known problem of their unambiguous definition \cite{dewitt,vil2}
does not arise in our consideration.
So, whether it is possible to extend the definition to higher energies
depends on eventual applicability of the classical conceptions contained
in the notion of measurement.

Now, turning back to the item 3. of the Introduction, it is natural to ask whether
the value of the effective gauge field, defined in the manner described above,
is one and the same for all measuring devices. It definitely is in the case
of infinitesimal device action, considered above. Indeed,
in this case account of any possible dependence of the effective gauge
field on characteristics of the measuring device would exceed the precision
chosen in our discussion. It is not clear, however, whether this is true
in the general case of finite disturbances produced in the effective
field by the process of measurement.

\section*{Acknowledgements}

We would like to thank our colleagues at the department of Theoretical Physics, Moscow State University, for many useful discussions.

We are also grateful to V.~V.~Asadov for substantial financial support of our research.

\end{document}